# Why is Astronomy Important?

By Marissa Rosenberg, Pedro Russo (EU-UNAWE, Leiden Observatory/Leiden University, The Netherlands / russo@strw.leidenuniv.nl), Georgia Bladon, Lars Lindberg Christensen (ESO, Germany)

## Introduction

Throughout History humans have looked to the sky to navigate the vast oceans, to decide when to plant their crops and to answer questions of where we came from and how we got here. It is a discipline that opens our eyes, gives context to our place in the Universe and that can reshape how we see the world. When Copernicus claimed that Earth was not the centre of the Universe, it triggered a revolution. A revolution through which religion, science, and society had to adapt to this new world view.

Astronomy has always had a significant impact on our world view. Early cultures identified celestial objects with the gods and took their movements across the sky as prophecies of what was to come. We would now call this astrology, far removed from the hard facts and expensive instruments of today's astronomy, but there are still hints of this history in modern astronomy. Take, for example, the names of the constellations: Andromeda, the chained maiden of Greek mythology, or Perseus, the demi-god who saved her.

Now, as our understanding of the world progresses, we find ourselves and our view of the world even more entwined with the stars. The discovery that the basic elements that we find in stars, and the gas and dust around them, are the same elements that make up our bodies has further deepened the connection between us and the cosmos. This connection touches our lives, and the awe it inspires is perhaps the reason that the beautiful images astronomy provides us with are so popular in today's culture.

There are still many unanswered questions in astronomy. Current research is struggling to understand questions like: "How old are we?", "What is the fate of the Universe?" and possibly the most interesting: "How unique is the Universe, and could a slightly different Universe ever have supported life?" But astronomy is also breaking new records every day, establishing the furthest distances, most massive objects, highest temperatures and most violent explosions.

Pursuing these questions is a fundamental part of being human, yet in today's world it has become increasingly important to be able to justify the pursuit of the answers. The difficulties in describing the importance of astronomy, and fundamental research in general, are well summarized by the following quote:

*"Preserving knowledge is easy. Transferring knowledge is also easy. But making new knowledge is neither easy nor profitable in the short term. Fundamental research proves profitable in the long run, and, as importantly, it is a force that enriches the culture of any society with reason and basic truth."*

- Ahmed Zewali, winner of the Nobel Prize in Chemistry (1999).

Although we live in a world faced with the many immediate problems of hunger, poverty, energy and global warming, we argue that astronomy has long term benefits that are equally as important to a civilized society. Several studies (see below) have told us that investing in science education, research and technology provides a great return — not only economically, but culturally and indirectly for the population in general — and has helped countries to face and overcome crises. The scientific and technological development of a country or region is closely linked to its human development index — a statistic that is a measure of life expectancy, education and income (Truman, 1949).

There are other works that have contributed to answering the question "Why is astronomy important?" Dr. Robert Aitken, director of Lick Observatory, shows us that even in 1933 there was a need to justify our science, in his paper entitled *The Use of Astronomy* (Aitken, 1933). His last sentence summarizes his sentiment: *"To give man ever more knowledge of the universe and to help him 'to learn humility and to know exaltation', that is the mission of astronomy."* More recently, C. Renée James wrote an article outlining the recent technological advances that we can thank astronomy for, such as GPS, medical imaging, and wireless internet (Renée James, 2012). In defence of radio astronomy, Dave Finley in Finley (2013) states, *"In sum, astronomy has been a cornerstone of technological progress throughout history, has much to contribute in the future, and offers all humans a fundamental sense of our place in an unimaginably vast and exciting universe."*

Astronomy and related fields are at the forefront of science and technology; answering fundamental questions and driving innovation. It is for this reason that the [International Astronomical Union](#)'s (IAU) strategic plan for 2010–2020 has three main areas of focus: technology and skills; science and research; and culture and society.

Although "blue-skies research" like astronomy rarely contributes directly with tangible outcomes on a short time scale, the pursuit of this research requires cutting-edge technology and methods that can on a longer time scale, through their broader application make a difference.

A wealth of examples — many of which are outlined below — show how the study of astronomy contributes to technology, economy and society by constantly pushing for instruments, processes and software that are beyond our current capabilities.

The fruits of scientific and technological development in astronomy, especially in areas such as optics and electronics, have become essential to our day-to-day life, with applications such as personal computers, communication satellites, mobile phones, [Global Positioning Systems](#), solar panels and [Magnetic Resonance Imaging](#) (MRI) scanners.

Although the study of astronomy has provided a wealth of tangible, monetary and technological gains, perhaps the most important aspect of astronomy is not one of economical measure. Astronomy has and continues to revolutionize our thinking on a worldwide scale. In the past, astronomy has been used to measure time, mark the seasons, and navigate the vast oceans. As one of the oldest sciences astronomy is part of every culture's history and roots. It inspires us with beautiful images and promises answers to the big questions. It acts as a window into the immense size and complexity of space, putting Earth into perspective and promoting global citizenship and pride in our home planet.

Several reports in the US (National Research Council, 2010) and Europe (Bode et al., 2008) indicate that the major contributions of astronomy are not just the technological and medical applications (technology transfer, see below), but a unique perspective that extends our horizons and helps us discover the grandeur of the Universe and our place within it. On a more pressing level, astronomy helps us study how to prolong the survival of our species. For example, it is critical to study the Sun's influence on Earth's climate and how it will affect weather, water levels etc. Only the study of the Sun and other stars can help us to understand these processes in their entirety. In addition, mapping the movement of all the objects in our Solar System, allows us to predict the potential threats to our planet from space. Such events could cause major changes to our world, as was clearly demonstrated by the meteorite impact in Chelyabinsk, Russia in 2013.

On a personal level, teaching astronomy to our youth is also of great value. It has been proven that pupils who engage in astronomy-related educational activities at a primary or secondary school are more likely to pursue careers in science and technology, and to keep up to date with scientific discoveries (National Research Council, 1991). This does not just benefit the field of astronomy, but reaches across other scientific disciplines.

Astronomy is one of the few scientific fields that interacts directly with society. Not only transcending borders, but actively promoting collaborations around the world. In the following paper, we outline the tangible aspects of what astronomy has contributed to various fields.

## Technology transfer

### From astronomy to industry

Some of the most useful examples of technology transfer between astronomy and industry include advances in imaging and communications. For example, a film called Kodak Technical Pan is used extensively by medical and industrial spectroscopists, industrial photographers, and artists, and was originally created so that solar astronomers could record the changes in the surface structure of the Sun. In addition, the development of Technical Pan — again driven by the requirements of astronomers — was used for several decades (until it was discontinued) to detect diseased crops and forests, in dentistry and medical diagnosis, and for probing layers of paintings to reveal forgeries (National Research Council, 1991).

In 2009 Willard S. Boyle and George E. Smith were awarded the Nobel Prize in Physics for the development of another device that would be widely used in industry. The sensors for image capture developed for astronomical images, known as Charge Coupled Devices (CCDs), were first used in astronomy in 1976. Within a very few years they had replaced film not only on telescopes, but also in many people's personal cameras, webcams and mobile phones. The improvement and popularity of CCDs is attributed to NASA's decision to use super-sensitive CCD technology on the [Hubble Space Telescope](.) (Kiger & English, 2011).

In the realm of communication, radio astronomy has provided a wealth of useful tools, devices, and data-processing methods. Many successful communications companies were originally founded by radio astronomers. The computer language [FORTH](.) was originally created to be used by the [Kitt Peak](.) 36-foot telescope and went on to provide the basis for a highly profitable company ([Forth Inc.](.)). It is now being used by FedEx worldwide for its tracking services.

Some other examples of technology transfer between astronomy and industry are listed below (National Research Council, 2010):

- The company General Motors uses the astronomy programming language [Interactive Data Language](.) (IDL) to analyse data from car crashes.
- The first patents for techniques to detect gravitational radiation — produced when massive bodies accelerate — have been acquired by a company to help them determine the gravitational stability of underground oil reservoirs.
- The telecommunications company [AT&T](.) uses [Image Reduction and Analysis Facility](.) (IRAF) — a collection of software written at the [National Optical Astronomy Observatory](.) — to analyse computer systems and solid-state physics graphics.
- Larry Altschuler, an astronomer, was responsible for the development of tomography - the process of imaging in sections using a penetrating wave - via his work on reconstructing the Solar Corona from its projections. (schuler, M. D. 1979)

### From astronomy to the aerospace sector

The aerospace sector shares most of its technology with astronomy — specifically in telescope and instrument hardware, imaging, and image-processing techniques.

Since the development of space-based telescopes, information acquisition for defence has shifted from using ground-based to aerial and space-based, techniques. Defence satellites are essentially telescopes pointed towards Earth and require identical technology and hardware to those used in their astronomical counterparts. In addition, processing satellite images uses the same software and processes as astronomical images.

Some specific examples of astronomical developments used in defence are given below (National Research Council, 2010):

- Observations of stars and models of stellar atmospheres are used to differentiate between rocket plumes and cosmic objects. The same method is now being studied for use in early warning systems.
- Observations of stellar distributions on the sky — which are used to point and calibrate telescopes — are also used in aerospace engineering.
- Astronomers developed a solar-blind photon counter — a device which can measure the particles of light from a source, during the day, without being overwhelmed by the particles coming from the Sun. This is now used to detect ultraviolet (UV) photons coming from the exhaust of a missile, allowing for a virtually false-alarm-free [UV](.)

[missile warning system](#). The same technology can also be used to detect toxic gases.
- Global Positioning System (GPS) satellites rely on astronomical objects, such as quasars and distant galaxies, to determine accurate positions.

**From astronomy to the energy sector**

Astronomical methods can be used to find new fossil fuels as well as to evaluate the possibility of new renewable energy sources (National Research Council, 2010):

- Two oil companies, [Texaco](#) and [BP](#), use IDL to analyse core samples around oil fields as well as for general petroleum research.
- An Australian company, called [Ingenero](#), has created solar radiation collectors to harness the power of the Sun for energy on Earth. They have created collectors up to 16 metres in diameter, which is only possible with the use of a graphite composite material developed for an orbiting telescope array.
- Technology designed to image X-rays in [X-ray telescopes](#) — which have to be designed differently from visible-light telescopes — is now used to monitor [plasma fusion](#). If fusion — where two light atomic nuclei fuse to form a heavier nucleus — became possible to control, it could be the answer to safe, clean, energy.

## Astronomy and medicine

Astronomers struggle constantly to see objects that are ever dimmer and further away. Medicine struggles with similar issues: to see things that are obscured within the human body. Both disciplines require high-resolution, accurate and detailed images. Perhaps the most notable example of knowledge transfer between these two studies is the technique of [aperture synthesis](#), developed by the radio astronomer and Nobel Laureate, Martin Ryle (Royal Swedish Academy of Sciences, 1974). This technology is used in [computerised tomography](#) (also known as CT or CAT scanners), [magnetic resonance imaging](#) (MRIs), [positron emission tomography](#) (PET) and many other medical imaging tools.

Along with these imaging techniques, astronomy has developed many programming languages that make image processing much easier, specifically IDL and IRAF. These languages are widely used for medical applications (Shasharina, 2005).

Another important example of how astronomical research has contributed to the medical world is in the development of clean working areas. The manufacture of space-based telescopes requires an extremely clean environment to prevent dust or particles that might obscure or obstruct the mirrors or instruments on the telescopes (such as in NASA's [STEREO mission](#); Gruman, 2011). The cleanroom protocols, air filters, and bunny suits that were developed to achieve this are now also used in hospitals and pharmaceutical labs (Clark, 2012).

Some more direct applications of astronomical tools in medicine are listed below:

- A collaboration between a drug company and the [Cambridge Automatic Plate Measuring Facility](#) allows blood samples from leukaemia patients to be analysed

- faster and thus ensures more accurate changes in medication (National Research Council, 1991).
- Radio astronomers developed a method that is now used as a non-invasive way to detect tumours. By combining this with other traditional methods, there is a true-positive detection rate of 96% in breast cancer patients (Barret et al., 1978).
- Small thermal sensors initially developed to control telescope instrument temperatures are now used to control heating in neonatology units — units for the care of newborn babies (National Research Council, 1991).
- A low-energy X-ray scanner developed by NASA is currently used for outpatient surgery, sports injuries, and in third-world clinics. It has also been used by the US Food and Drugs Administration (FDA) to study whether certain pills had been contaminated (National Research Council, 1991).
- Software for processing satellite pictures taken from space is now helping medical researchers to establish a simple method to implement wide-scale screening for Alzheimer's disease (ESA, 2013).
- Looking through the fluid-filled, constantly moving eye of a living person is not that different from trying to observe astronomical objects through the turbulent atmosphere, and the same fundamental approach seems to work for both. Adaptive optics used in astronomy can be used for retinal imaging in living patients to study diseases such as macular degeneration and retinitis pigmentosa in their early stages. (Boston Micromachines Corporation 2010)

## Astronomy in everyday life

There are many things that people encounter on an everyday basis that were derived from astronomical technologies. Perhaps the most commonly used astronomy-derived invention is the wireless local area network (WLAN). In 1977 John O'Sullivan developed a method to sharpen images from a radio telescope. This same method was applied to radio signals in general, specifically to those dedicated to strengthening computer networks, which is now an integral part of all WLAN implementations (Hamaker et al., 1977).

Other technologies important to everyday life that were originally developed for astronomy are listed below (National Research Council, 2010):

- X-ray observatory technology is also used in current X-ray luggage belts in airports.
- In airports, a gas chromatograph — for separating and analysing compounds — designed for a Mars mission is used to survey baggage for drugs and explosives.
- The police use hand-held Chemical Oxygen Demand (COD) photometers — instruments developed by astronomers for measuring light intensity — to check that car windows are transparent, as determined by the law.
- A gamma-ray spectrometer originally used to analyse lunar soil is now used as a non-invasive way to probe structural weakening of historical buildings or to look behind fragile mosaics, such as in St. Mark's Basilica in Venice.

More subtle than these contributions to technology is the contribution that astronomy has made to our view of time. The first calendars were based on the movement of the Moon and even the way that we define a second is due to astronomy. The atomic clock, developed in

1955, was calibrated using astronomical Ephemeris Time — a former standard astronomical timescale adopted by the IAU in 1952. This led to the internationally agreed-upon re-definition of the second (Markowitz et al., 1958).

These are all very tangible examples of the effect astronomy has had on our everyday lives, but astronomy also plays an important role in our culture. There are many books and magazines about astronomy for non-astronomers. [A Brief History of Time](#) by Stephen Hawking is a bestseller and has sold over ten million copies (Paris, 2007) and Carl Sagan's television series, [Cosmos: A Personal Voyage](#), has been watched in over 60 countries by more than 500 million people (NASA, 2009).

Many non-astronomers also engaged with astronomy during the [International Year of Astronomy 2009](#) (IYA2009), the largest education and public outreach event in science. The IYA2009 reached upwards of eight hundred million people, through thousands of activities, in more than 148 countries (IAU, 2010).

## Astronomy and international collaboration

Scientific and technological achievements give a large competitive edge to any nation. Nations pride themselves on having the most efficient new technologies and race to achieve new scientific discoveries. But perhaps more important is the way that science can bring nations together, encouraging collaboration and creating a constant flow as researchers travel around the globe to work in international facilities.

Astronomy is particularly well suited to international collaboration due to the need to have telescopes in different places around the world, in order to see the whole sky. At least as far back as 1887 — when astronomers from around the world pooled their telescope images and made the first map of the whole sky — there have been international collaborations in astronomy and in 1920, the International Astronomical Union became the first international scientific union.

In addition to the need to see the sky from different vantage points on Earth, building astronomical observatories on the ground and in space is extremely expensive. Therefore most of the current and planned observatories are owned by several nations. All of these collaborations have thus far been peaceful and successful. Some of the most notable being:

- The [Atacama Large Millimeter/submillimeter Array (ALMA)](#), an international partnership of Europe, North America and East Asia in cooperation with the Republic of Chile, is the largest astronomical project in existence.
- The [European Southern Observatory](#) (ESO) which includes 14 European countries and Brazil, and is located in Chile.
- Collaborations on major observatories such as the NASA/ESA Hubble Space Telescope between USA and Europe.

## Summary

In the above text we have outlined both the tangible and intangible reasons that astronomy is an important part of society.  Although we have focused mainly on the technology and knowledge transfer, perhaps the most important contribution is still the fact that astronomy makes us aware of how we fit into the vast Universe.  The American astronomer Carl Sagan showed us one of astronomy's simplest and most inspirational contributions to society in his book, [The Pale Blue Dot](#):

*"It has been said that astronomy is a humbling and character-building experience. There is perhaps no better demonstration of the folly of human conceits than this distant image of our tiny world. To me, it underscores our responsibility to deal more kindly with one another, and to preserve and cherish the pale blue dot, the only home we've ever known."*